\documentclass[conference]{IEEEtran}
\ifCLASSINFOpdf
\else
\fi
\usepackage{amsfonts}
\usepackage{amssymb}
\usepackage{amsmath,bm}
\usepackage{graphicx}
\usepackage{algorithm}
\usepackage{algorithmic}
\usepackage{multirow}
\usepackage{amsmath}
\usepackage{xcolor}
\usepackage{subfigure}
\usepackage{url}

\begin{document}
%
\title{Identifying Talented Software Engineering Students through Data-driven Skill Assessment}



%
\author{
\IEEEauthorblockN{Jun Lin\IEEEauthorrefmark{1}\IEEEauthorrefmark{2},
Han Yu\IEEEauthorrefmark{1} and
Zhiqi Shen\IEEEauthorrefmark{1}}

\IEEEauthorblockA{\IEEEauthorrefmark{1}School of Computer Engineering, Nanyang Technological University, Singapore}
\IEEEauthorblockA{\IEEEauthorrefmark{2}School of Software, Beihang University, Beijing, China}

\IEEEauthorblockN{\{jlin7, han.yu, zqshen\}@ntu.edu.sg, linjun@buaa.edu.cn}
}


\maketitle

\begin{abstract}
For software development companies, one of the most important objectives is to identify and acquire talented software engineers in order to maintain a skilled team that can produce competitive products. Traditional approaches for finding talented young software engineers are mainly through programming contests of various forms which mostly test participants' programming skills. However, successful software engineering in practice requires a wider range of skills from team members including analysis, design, programming, testing, communication, collaboration, and self-management, etc. In this paper, we explore potential ways to identify talented software engineering students in a data-driven manner through an Agile Project Management (APM) platform. Through our proposed \textit{HASE} online APM tool, we conducted a study involving 21 Scrum teams consisting of over 100 undergraduate software engineering students in multi-week coursework projects in 2014. During this study, students performed over 10,000 ASD activities logged by HASE. We demonstrate the possibility and potentials of this new research direction, and discuss its implications for software engineering education and industry recruitment.
\end{abstract}


%
\IEEEpeerreviewmaketitle

\section{Introduction}
Most hiring managers in software engineering companies understand that a successful member of software development team need to be strong in both \textit{programming skills} (e.g., software design, coding, and testing skills) and \textit{soft skills} (e.g., communication, collaboration, and self-management skills) \cite{Freeman-et-al:1976}. Programming skills can be gauged, at least in part, from students' performance in examinations and programming contests. Soft skills are much harder to assess, especially during the limited time given in job interviews. Although the concept of these skills can be taught in classroom, the ability to apply them consistently in practice can only be acquired through one's own experience.

As today's software engineers often need to work in a team environment when developing complex software systems, hiring one who works well with other is an important consideration. Interviews have been specifically designed for this purpose. By listening to what a candidate says and observing what he does, an experienced hiring manager is able to determine whether this person has the right skill set for the job. However, through such short interactions and with limited quantifiable data, it is challenging for even a highly experience hiring manager to make these judgements.

With the emergence of systems capable of collecting personal behavior trajectory big data \cite{Heymann-Garcia-Molina:2011}, data-driven analysis of people's characteristics over time is bringing a revolution in how students' learning performance can be measured. For example, the Ministry of Education in Singapore has started an initiative to build technological solutions capable of holistically assessing students' 21st Century Competencies (e.g., critical thinking, self-directed learning skills) in recent years (\url{http://www.moe.gov.sg/media/press/2010/03/moe-to-enhance-learning-of-21s.php}). Preliminary works in this field using virtual world-based learning environments have already started \cite{Michael-et-al:2010}.

Following a similar line of thinking, we explore how software development behavior data can be used to assess students' programming skills and soft skills.
Since Agile Software Development (ASD) involves more human factors reflecting developers' personal characteristics compared to other plan-driven methodologies \cite{Cockburn-Highsmith:2011}, we focus on tracking students' development activities and behaves in the ASD process. For this purpose, we conduct a 12 week study involving 125 undergraduate software engineering students. The students self-organized into 21 ASD teams of 5 to 7 persons. Each team developed one software system following the Scrum ASD methodology as part of their coursework requirements.

Students in this study carried out software engineering activities at various stages of the Scrum methodology in our online agile project management (APM) tool - the Human-centred Agile Software Engineering (HASE) platform\footnote{\url{http://www.linjun.net.cn/hase/}} \cite{Lin-et-al:2014}. The activities for each team member supported by HASE mainly occur during the sprint planning and sprint review/retrospective phases. They include proposing tasks, estimating the priority, difficulty and time required for each task, deciding how to allocate tasks, collaboration information, reviewing the timeliness and quality of completed tasks, and providing feedback on individual team member's mood at different points in time during a sprint. During the study, students logged 10,779 ASD activities in the HASE platform. By analyzing the collected to reflect students' programming skills, effectiveness of collaboration, and emotional stability, we demonstrate the potential of this research direction and discuss its implications for software engineering education and industry recruitment.

\section{Related Work} \label{st:related-work}
To the best of our knowledge, there has yet to be published previous studies using software development behavior data to assess software engineers skills. Nevertheless, as the skills assessment has always been an important problem, other methods have been applied in an attempt to address it.

In \cite{Salleh-et-al:2011}, the authors present the results of a systematic literature review concerning agile pair programming effectiveness. The paper analyzed compatibility factors, such as the feel good, personality, and skill level factors, and their effect on pair programming effectiveness as a pedagogical tool in Computer Science and Software Engineering education. Four metrics were used in the analysis: 1) academic performance, 2) technical productivity, 3) program/design quality and 4) learning satisfaction. As the study was not focused on assessment, the general findings are that pair programming is more effective in terms of technical productivity, learning satisfaction and academic performance, while not significantly different in terms of program quality as compared to solo programming. Nevertheless, it did point towards the importance of soft skills in successful software engineering.

The study reported in \cite{Lin-et-al:2014} started to track personal performance data with agile project management tools to study task allocation related decision-making under Scrum. It employed the same research techniques as reported in this paper. However, the study in \cite{Lin-et-al:2014} focused on analyzing students' programming skills and did not consider their soft skills such as collaboration and morale.

\begin{figure*}
    \centering
    \subfigure[Students' competence v.s. productivity]{
    \includegraphics[trim = 40mm 85mm 45mm 90mm, clip, width = 2.2in]{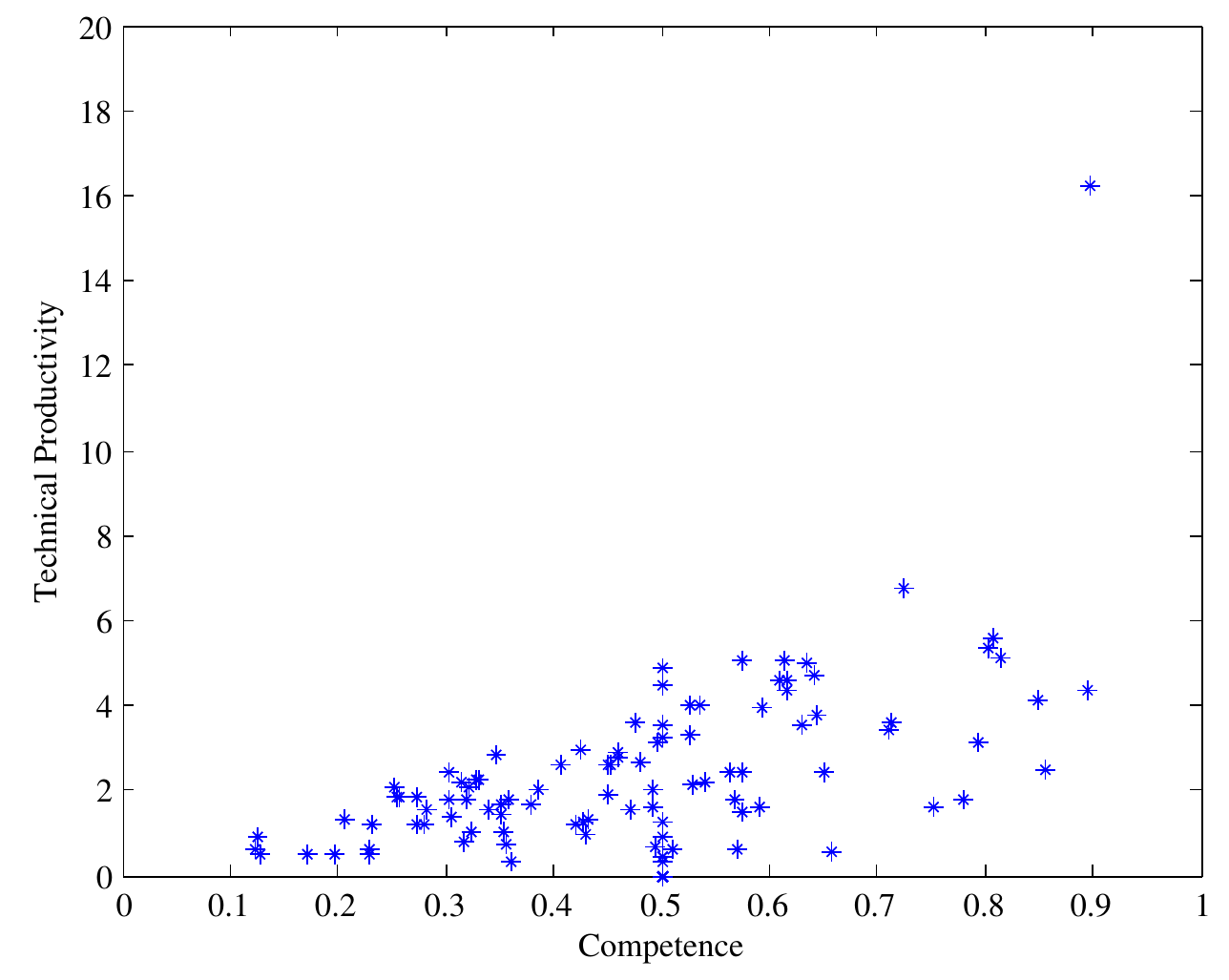}\label{fig:StudentsCompCap}}
    \subfigure[Average number of collaborators per task]{
    \includegraphics[trim = 40mm 85mm 45mm 90mm, clip, width = 2.2in]{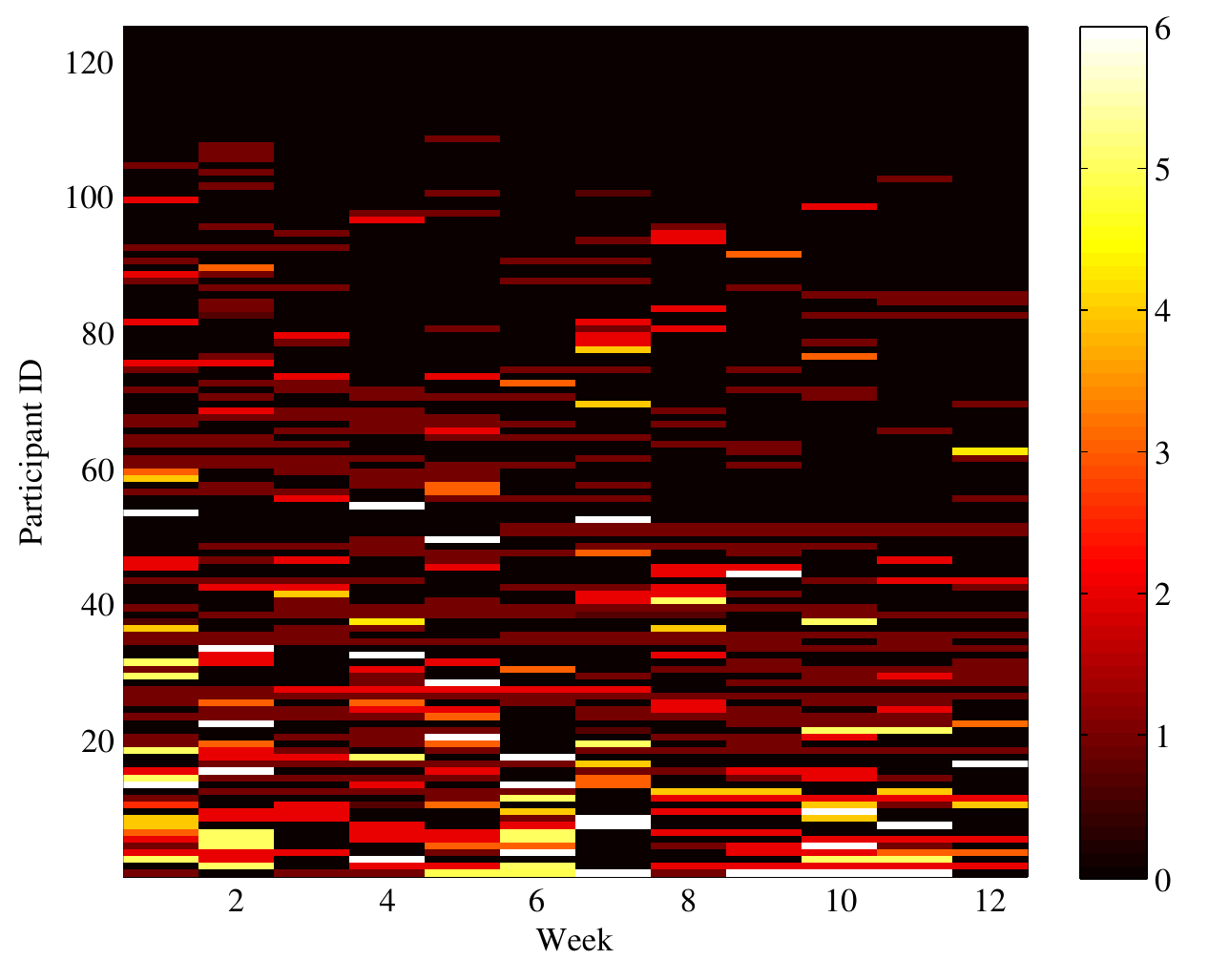}\label{fig:Collaborations}}
    \subfigure[Intra-week mood variation]{
    \includegraphics[trim = 40mm 85mm 45mm 90mm, clip, width = 2.2in]{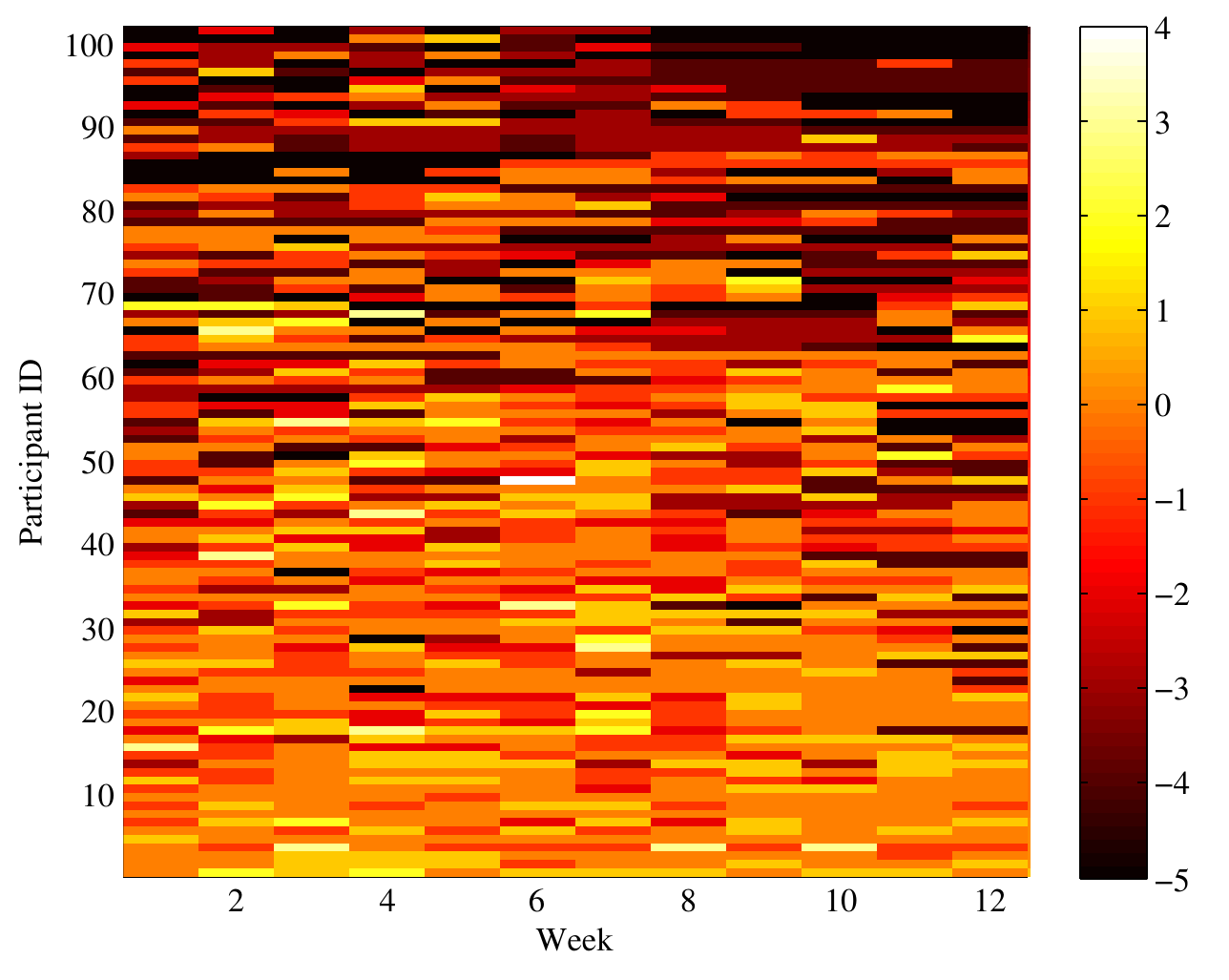}\label{fig:IntraWeek}}
    \vspace*{-5pt}
    \caption{Participants' Characteristics.} 
    \vspace*{-5pt}
\end{figure*}

\section{Study Design} \label{st:study-design}
In this section, we present our research approach and the metrics that have been adopted to measure students' skills in our study.

\subsection{Research Approach}
Our goal is to investigate the activities and behaviors of students, including their decision-making, collaboration, task assignment and mood etc. in the Scrum ASD process. The platform provides five main features to support agile project management which cover the sprint planning and sprint review/retrospective phases:
\begin{enumerate}
  \item \textit{Registration}: In order to build user profiles, HASE requires registrants to specify their self-assessed competence levels in different areas of expertise such as familiarity with specific programming languages, system design methodologies, and user interface (UI) design tools, etc.
  \item \textit{Team and Role Management}: HASE supports the creation of teams, the selection of product owners and stakeholders into the teams, and the assignment of different roles within a team (e.g., programmers and UI designers).
  \item \textit{Task Management}: Task information including task description, skills required for the task, and the person who proposed each task is displayed for all team members to view. The difficulty value of each task $\tau$, is recorded using an 11-point Likert scale \cite{Likert:1932} (with 0 denoting ``extremely easy" and 10 denoting ``extremely hard"). Each team member can input his/her estimated difficulty value for each task into the HASE platform. The HASE platform then uses the average difficulty value for the task ($D_{\tau}$). The students were asked to take into account the technical challenge as well as the amount of effort required when judging the difficulty of a task. The priority value of each task $\tau$, is also recorded using an 11-point Likert scale (with 0 denoting ``extremely low priority" and 10 denoting ``extremely high priority"). Each team member can input his/her estimated priority value for each task into the HASE platform. The HASE platform then uses the average priority value for the task.
  \item \textit{Sprint Planning}: HASE records the teams' decisions on which tasks are assigned to which team member during each sprint. Once assigned, the status of the task becomes ``Assigned". The assignee $i$ inputs his/her confidence value ($Conf_{\tau}^{i}$) for each task $\tau$ on an 11-point Likert scale (with 0 denoting ``not confident at all" and 10 denoting ``extremely confident"). Each team member also inputs the estimated required time to complete each task (in number of days). The HASE platform uses the average estimated time required to generate the deadline for the task ($T_{\tau}^{est}$). Apart from a primary assignee, multiple students can collaboratively work on a task. The collaborator information for each task is also recorded by HASE.
  \item \textit{Sprint Review/Retrospective}: Once a task is completed, the assignee changes its status in the HASE platform to ``Completed". This action will trigger HASE to record the actual number of days ($T_{\tau}^{act}$) used to complete this task. HASE also provides functions for team members to peer review the quality ($Qual_{\tau}$) of each completed task $\tau$. The quality of a completed task is recorded in the platform using a 11-point Likert scale with 0 representing (``extremely low quality") and 10 representing (``extremely high quality"). The average quality rating for each task is used by HASE as the final quality rating for that task.
  \item \textit{Team Morale Monitoring}: During the sprint planning meeting, team members can report their current mood values into the HASE platform. A person $i$'s mood at the beginning of a sprint $t$ ($m_{i}^{begin}(t)$) is represented on a 5-point Likert scale with 1 representing ``very low" and 5 representing ``very high". During the sprint review/retrospective meeting, each task assignee $i$ can report his/her mood after completing a task at the end of sprint $t$ ($m_{i}^{end}(t)$) using the same 5-point Likert scale.
\end{enumerate}

\subsection{Metrics}
The following metrics have been adopted to facilitate our analysis:

\textit{Technical Productivity} ($\mu_{i}$): it refers to the average amount of workload a student $i$ can complete during a sprint. In this study, we use the task difficulty value as an indicator of the workload of a task as the task difficulty values reported by students denote both the technical challenge and the amount of effort required to complete the task.

\textit{Competence} ($Comp_{i}$): it refers to the probability a student $i$ can complete a task assigned to him/her with satisfactory quality before the stipulated deadline. In this paper, the outcome of a task needs to achieve an average quality rating higher than 5 out of 10 in order to be considered as having satisfactory quality. This metric is similar to a student's reputation. Thus, we adopt a reputation computation model - the Beta Reputation model \cite{Josang-et-al:2007} - which is widely used in the fields of online services, artificial intelligence and network communications \cite{Pan-et-al:2009,Yu-et-al:2010,Yu-et-al:2011,Yu-et-al:2013}. It is calculated as follows:
\begin{equation}
Comp_{i}=\frac{\alpha_{i}+1}{(\alpha_{i}+1)+(\beta_{i}+1)}\in (0,1) \label{eq:1}
\end{equation}
where $\alpha_{i}$ and $\beta_{i}$ are calculated as:
\begin{equation}
    \alpha_{i}= \sum_{\tau\in \phi(i)}1_{[T_{\tau}^{act}-T_{\tau}^{est}\leq 0 \mbox{ and } Qual_{\tau}>5]}D_{\tau} \label{eq:2}
\end{equation}
\begin{equation}
    \beta_{i}= \sum_{\tau\in \phi(i)}1_{[T_{\tau}^{act}-T_{\tau}^{est}>0 \mbox{ or } Qual_{\tau}\leq 5]}D_{\tau}. \label{eq:3}
\end{equation}
The function $1_{[\mbox{condition}]}$ in Eq. (\ref{eq:2}) and Eq. (\ref{eq:3}) equals to 1 if ``condition" is true. Otherwise, $1_{[\mbox{condition}]}$ equals to 0. $\phi(i)$ denotes the set of tasks $i$ has previously worked on until the current point in time. The ``+1" terms in the numerator and denominator of Eq. (\ref{eq:1}) ensure that if $i$ has no previous track record, $Comp_{i}$ evaluates to 0.5 indicating maximum uncertainty about $i$'s performance.

\section{Results and Analysis} \label{st:results}

An initial exploratory data analysis has identified certain personal characteristics which may become useful markers for assessing students' skills in the future. Figure \ref{fig:StudentsCompCap} shows the participants' competence scores versus their productivity scores at the end of the study. It can be observed that the participants' performance according to these two metrics is quite diverse. In general, participants who demonstrated high competence tend to also be able to handle high workloads allocated to them ($r=0.7443$, $p<0.01$). One participant achieved significantly higher competence and productivity score than the rest of the participants.

Collaboration is generally regarded as a useful way to improve the effectiveness and efficiency of a software team. Figure \ref{fig:Collaborations} shows a heat map of the number of collaborators per task each participant had for each of the 12 weeks. The lighter the color of a point on the figure, the more collaborators per task that participant had for that particular week. The color scale mapping different color gradients to the actual number of collaborators per task is shown on the right-hand side of the figure. Participants are ranked in descending order of their average number of collaborators per task per week. Those who are shown at the bottom of the figure ranked the highest among their peers. It can be observed that this metric can distinguish the behavior among different participants clearly.

Stability of mood is a sign showing one's maturity and self-management skills. Figure \ref{fig:IntraWeek} shows a heat map of the intra week mood change (which is computed as $\Delta m_{i}(t)=m_{i}^{end}(t)-m_{i}^{begin}(t)\in [-5,5]$ for each week) over the 12 weeks. The color scale mapping different color gradients to the intra week mood change is shown on the right-hand side of the figure. Participants are ranked in descending order of their intra week mood change values per week. Those who are shown at the bottom of the figure ranked the highest among their peers. It can be observed that this metric can distinguish the behavior among different participants quite clearly. The mood of those who ranked high on this metric tends to increase within at the end of a week after a sprint of development. And as their mood at the beginning of the week also tend to be high, the increments are generally small. Thus, their mood remain relatively stable throughout a sprint. Those who ranked low on this metric (top part of the figure) tend to have big negative mood swings, especially towards the end of the study.

In order to explore if the assessment of participants' skills may help us identify students who are good at hands-on software engineering but did not stand out in examinations, we construct a \textit{skills score} to aggregate the effect of competence, productivity, collaboration, and mood stability into one scalar measurement. In this study, the skills score, $S_{skills}(i)$, for a participant $i$ is computed as:
\begin{equation}
S_{skills}=\frac{S_{\mu_{i}}+S_{Comp_{i}}+S_{col_{i}}}{5-S_{\Delta m_{i}}},
\end{equation}
where $S_{\mu_{i}}$, $S_{Comp_{i}}$, $S_{col_{i}}$ and $S_{\Delta m_{i}}$ are the normalized scores for $i$ in terms of productivity, competence, collaboration, and mood stability, respectively ($S_{skills}(i)\in[0,100]$).

\begin{figure}
    \centering
    \includegraphics[trim = 40mm 85mm 45mm 90mm, clip, width = 3.3in]{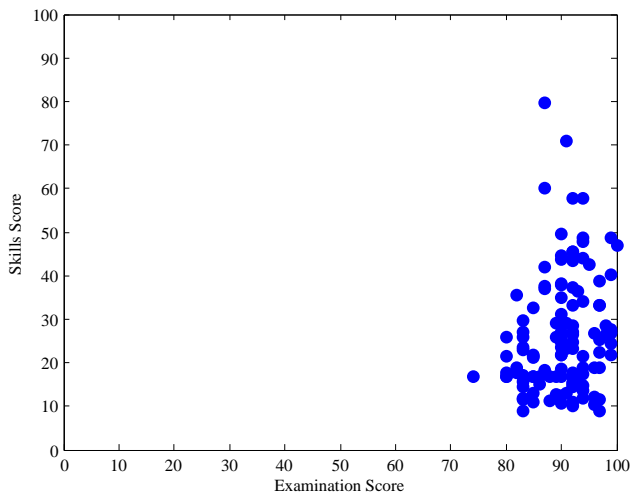}
    \caption{Skills score v.s. examination score.} \label{fig:TwoScores}
\end{figure}

Figure \ref{fig:TwoScores} plots the participants' skills score against their examination score for the subject of software engineering in the same semester. It can be observed that, according to their examination score, their performance clustered in the range of 80 to 100 marks, enabling almost all of them to achieve a grade of A or A+. However, their skills scores spread from as low as 10 marks to as high as 80 marks, making their performance more distinguishable compared to their examination scores. Furthermore, the top three best performing participants in terms of skills scores achieved only average scores in their examinations. We acknowledge that there may be other ways to compute the skills score and we refrain from claiming that our current formulation for the skills score is the most effective. Nevertheless, the results show that the data-driven skills score can indeed help us identify talented software engineering students whom examinations based assessments failed to identify.

\section{Discussions and Future Work} \label{st:conclusions}
In this paper, we explore a novel data-driven Approach to find talented software engineering students using an Agile Project Management Platform. Different from traditional interview/internship-based methods, our study is based on participants' ASD activity trajectory data collected unobtrusively during normal ASD processes through our HASE APM platform. This type of data objectively reflects developers' ASD activities and performance at fine granularity.

As the data collection and analytics technologies further develop, software engineering students may eventually perform all coursework activities in a technology platform capable of unobtrusively collecting their behavior data and continuously assessing a wide range of their skills over time. In this way, the students' practical skill development can be monitored by their instructors so that they can adjust their pedagogical methods and give personalized attention to students who need more help in particular areas. Such a tool will enable software engineering educators to have a quantifiable way of understanding their students' skill development and take a proactive approach in helping them develop programming and soft skills. The skills scores may, one day, be taken into consideration by industry recruiters as a complement to a candidate's academic profile to help companies identify well rounded software engineering talents suitable for their teams.

With this study, we see the start of a series of research and application on agile software development with ASD activity trajectory data, especially for recommending of talented students to industry. In future research, we will explore various modeling approaches (e.g., fuzzy cognitive approaches \cite{Miao-et-al:2002,Song-et-al:2009}, evolutionary methods \cite{Li-et-al:2009}, and inference models \cite{Miao-et-al:2001}) to design personalized inference models to convert the behavior trajectory data into predictive analytics models to help instructors gain deeper understanding into students' skills development. We plan to conduct surveys/interviews to understand more in-depth how students in each Scrum team collaborate. We will continue using the HASE platform to collect agile programming activity data over subsequent semesters and expand our data collection effort to include more universities so as to investigate the possible effects of socio-cultural factors. More finely grained data such as the time each student spent on a task and the breakdown of the usage of the time (e.g., how much time is spent on reading task requirements, designing, discussions, and coding) will also be collected in future versions of the HASE platform. The resulting datasets will be published in the future to support the discovery of new insights by researchers in the field.

\section*{Acknowledgment}
This research is supported by the  National Research Foundation, Prime Minister's Office, Singapore under its IDM Futures Funding Initiative and administered by the Interactive and Digital Media Programme Office.



%

\bibliographystyle{IEEEtran}
\bibliography{Reference}

\begin{thebibliography}{10}
\providecommand{\url}[1]{#1}
\csname url@samestyle\endcsname
\providecommand{\newblock}{\relax}
\providecommand{\bibinfo}[2]{#2}
\providecommand{\BIBentrySTDinterwordspacing}{\spaceskip=0pt\relax}
\providecommand{\BIBentryALTinterwordstretchfactor}{4}
\providecommand{\BIBentryALTinterwordspacing}{\spaceskip=\fontdimen2\font plus
\BIBentryALTinterwordstretchfactor\fontdimen3\font minus
  \fontdimen4\font\relax}
\providecommand{\BIBforeignlanguage}[2]{{%
\expandafter\ifx\csname l@#1\endcsname\relax
\typeout{** WARNING: IEEEtran.bst: No hyphenation pattern has been}%
\typeout{** loaded for the language `#1'. Using the pattern for}%
\typeout{** the default language instead.}%
\else
\language=\csname l@#1\endcsname
\fi
#2}}
\providecommand{\BIBdecl}{\relax}
\BIBdecl

\bibitem{Freeman-et-al:1976}
P.~Freeman, A.~I. Wasserman, and R.~E. Fairley, ``Essential elements of
  software engineering education,'' in \emph{Proceedings of the 2Nd
  International Conference on Software Engineering (ICSE'76)}, 1976, pp.
  116--122.

\bibitem{Heymann-Garcia-Molina:2011}
P.~Heymann and H.~Garcia-Molina, ``Turkalytics: analytics for human
  computation,'' in \emph{Proceedings of the 20th International Conference on
  World Wide Web (WWW'11)}, 2011, pp. 477--486.

\bibitem{Michael-et-al:2010}
M.~J. Jacobson, B.~Kim, C.~Miao, Z.~Shen, and M.~Chavez, ``Design perspectives
  for learning in virtual worlds,'' in \emph{Designs for Learning Environments
  of the Future}, M.~J. Jacobson and P.~Reimann, Eds.\hskip 1em plus 0.5em
  minus 0.4em\relax Springer US, 2010, pp. 111--141.

\bibitem{Cockburn-Highsmith:2011}
A.~Cockburn and J.~Highsmith, ``Agile software development, the people
  factor,'' \emph{Computer}, vol.~34, no.~11, pp. 131--133, 2001.

\bibitem{Lin-et-al:2014}
J.~Lin, H.~Yu, Z.~Shen, and C.~Miao, ``Studying task allocation decisions of
  novice agile teams with data from agile project management tools,'' in
  \emph{Proceedings of the 29th IEEE/ACM International Conference on Automated
  Software Engineering (ASE'14)}, 2014, pp. 689--694.

\bibitem{Salleh-et-al:2011}
N.~Salleh, E.~Mendes, and J.~Grundy, ``Empirical studies of pair programming
  for cs/se teaching in higher education: A systematic literature review,''
  \emph{IEEE Transactions on Software Engineering (TSE)}, vol.~37, no.~4, pp.
  509--525, 2011.

\bibitem{Likert:1932}
R.~Likert, ``A technique for the measurement of attitudes,'' \emph{Archives of
  Psychology}, vol.~22, no. 140, 1932.

\bibitem{Josang-et-al:2007}
A.~J{\o}sang, R.~Ismail, and C.~Boyd, ``A survey of trust and reputation
  systems for online service provision,'' \emph{Decision Support Systems
  (DSS)}, vol.~43, no.~2, pp. 618--644, 2007.

\bibitem{Pan-et-al:2009}
L.~Pan, X.~Meng, Z.~Shen, and H.~Yu, ``A reputation pattern for service
  oriented computing,'' in \emph{Proceedings of the 7th International
  Conference on Information, Communications and Signal Processing (ICICS'09)},
  2009.

\bibitem{Yu-et-al:2010}
H.~Yu, Z.~Shen, C.~Miao, C.~Leung, and D.~Niyato, ``A survey of trust and
  reputation management systems in wireless communications,'' \emph{Proceedings
  of the IEEE}, vol.~98, no.~10, pp. 1755--1772, 2010.

\bibitem{Yu-et-al:2011}
H.~Yu, S.~Liu, A.~C. Kot, C.~Miao, and C.~Leung, ``Dynamic witness selection
  for trustworthy distributed cooperative sensing in cognitive radio
  networks,'' in \emph{Proceedings of the 13th IEEE International Conference on
  Communication Technology (ICCT'11)}, 2011, pp. 1--6.

\bibitem{Yu-et-al:2013}
H.~Yu, Z.~Shen, C.~Leung, C.~Miao, and V.~R. Lesser, ``A survey of multi-agent
  trust management systems,'' \emph{IEEE Access}, vol.~1, no.~1, pp. 35--50,
  2013.

\bibitem{Miao-et-al:2002}
C.~Miao, Q.~Yang, H.~Fang, and A.~Goh, ``Fuzzy cognitive agents for
  personalized recommendation,'' in \emph{Proceedings of the 3rd International
  Conference on Web Information Systems Engineering (WISE'02)}, 2002, pp.
  362--371.

\bibitem{Song-et-al:2009}
H.~Song, C.~Miao, Z.~Shen, Y.~Miao, and B.-S. Lee, ``A fuzzy neural network
  with fuzzy impact grades,'' \emph{Neurocomputing}, vol.~72, no.~13, pp.
  3098--3122, 2009.

\bibitem{Li-et-al:2009}
B.~Li, H.~Yu, Z.~Shen, and C.~Miao, ``Evolutionary organizational search,'' in
  \emph{Proceedings of the 8th International Conference on Autonomous Agents
  and Multiagent Systems (AAMAS'09)}, 2009, pp. 1329--1330.

\bibitem{Miao-et-al:2001}
C.~Miao, A.~Goh, Y.~Miao, and Z.~Yang, ``A dynamic inference model for
  intelligent agents,'' \emph{International Journal of Software Engineering and
  Knowledge Engineering}, vol.~11, no.~05, pp. 509--528, 2001.

\end{thebibliography}

\end{document}